\documentclass[aps,amsmath,pra,showpacs,twocolumn,superscriptaddress]{revtex4}

\usepackage{graphics}

\newcommand \beq{\begin{equation}}
\newcommand \beqa{\begin{eqnarray}}
\newcommand \beqann{\begin{eqnarray*}}
\newcommand \eeq{\end{equation}}
\newcommand \eeqa{\end{eqnarray}}
\newcommand \eeqann{\end{eqnarray*}}

\begin{document}

\title{Efficient creation of maximally entangled states by modulation of tunneling rates
}

\author{Gentaro Watanabe}
\affiliation{RIKEN, 2-1 Hirosawa, Wako,
Saitama 351-0198, Japan}
\affiliation{CNR INFM-BEC and Department of Physics, University of Trento, I-38050 Povo, Italy}

\date{\today}

\begin{abstract}
For systems described by the two-site Bose-Hubbard Hamiltonian, we
show that a sinusoidal modulation of the tunneling matrix element
assists higher-order co-tunneling processes.  Using this mechanism, we
propose an efficient new scheme for creating a coherent superposition
of states in which all particles are either on one site or all on the
other site, the so-called NOON state.  This scheme yields an almost
perfect NOON state periodically.  For larger numbers of particles,
further reduction of the time to create the state is possible if more
than one modulation frequency is employed.  With this scheme, NOON
states with a larger number of particles could be realized with
state-of-the-art techniques for cold Bose gases in a double-well
potential.
\end{abstract}


\maketitle

%\section{Introduction}

One of the current challenges in quantum control is to create
particular sorts of entangled states that are a superposition of
states that are maximally different.  The largest number of particles
for which such states can be engineered is six for ions in traps
\cite{leibfried} and $\approx 10$ photons \cite{ourjoumtsev} for
photons.  Atoms in a double-well potential are a fundamental system
for quantum-state engineering and it is a promising candidate for
creating such superposition states.  Especially the coherent
superposition of a state in which all $N$ particles are in the right
well and that in which all particles are in the left well (a so-called
NOON state \cite{lee}) is a maximally entangled state
\cite{gisin,mermin}, and it is an important resource for matter-wave
interferometry: it provides the ultimate quantum limit of the phase
resolution $\sim 1/N$, i.e., the Heisenberg limit, for any $N$
\cite{gerry} (see also \cite{boto_pezze}).  However, NOON states with
larger $N$ are fragile with respect to decoherence and thus it is
crucial to create them in a short time.

In Ref.\ \cite{vavtunnel}, we studied the tunneling of bosons in a 
double-well potential \cite{note_map} described by the two-site
Bose-Hubbard Hamiltonian \cite{note_anharm}:
\begin{equation}
\hat{H}=-J ({\hat c}_R^\dagger {\hat c}_L + {\hat c}_L^\dagger {\hat c}_R)
+\frac{U}{2}({\hat c}_R^\dagger {\hat c}_R^\dagger {\hat c}_R {\hat c}_R
+{\hat c}_L^\dagger {\hat c}_L^\dagger {\hat c}_L {\hat c}_L) ,
\label{bh}
\end{equation}
where ${\hat c}_R^\dagger$ and ${\hat c}_L^\dagger$ create bosons in
the right and left well, respectively, $J$ is the tunneling matrix
element, and $U$ is the on-site interaction.  There, we found that,
starting from a situation in which all particles are in the right or
the left well, a NOON state is formed by co-tunneling provided the
interaction is strong enough: $\kappa/2 \equiv UN/2J \gg 1$.  In this
situation, single-particle tunneling is suppressed by the energy
conservation and all $N$ particles are forced to be in the same well.
While the classical theory predicts self-trapping
\cite{milburn,smerzi_st}, a superposition of all particles in the
right and left wells is allowed in quantum mechanics and an
oscillation between these two states occurs due to higher-order
co-tunneling, which yields a NOON state after a quarter of the
oscillation period \cite{vavtunnel}.  Indeed, a NOON state for $N=2$
would have been realized by this mechanism in experiments of Ref.\
\cite{mainzexp}.  A great advantage of this mechanism is that almost
{\it perfect} NOON states are obtained at periodic intervals in
double-well potentials unlike, e.g., the well-known protocol of Ref.\
\cite{gordon} (see also \cite{micheli,mahmud} and references therein;
protocols for different setups are described in, e.g., Refs.\
\cite{molmer,piazza}).  On the other hand, a serious difficulty with
the mechanism of Ref.\ \cite{vavtunnel} is that the time to form NOON
states is much larger than for other protocols and increases
exponentially with $N$.  In the present work, we shall show how this
drawback may be overcome.  We also show that, using our present
scheme, NOON states with $N$ comparable to the largest value attained
for trapped ions \cite{leibfried} can be realized with cold atoms
under state-of-the-art experimental conditions \cite{mainzexp}.

%\section{New scheme for creating the NOON state}

\begin{figure}[tbp]
%\begin{center}
\rotatebox{0}{
\resizebox{8.2cm}{!}
{\includegraphics{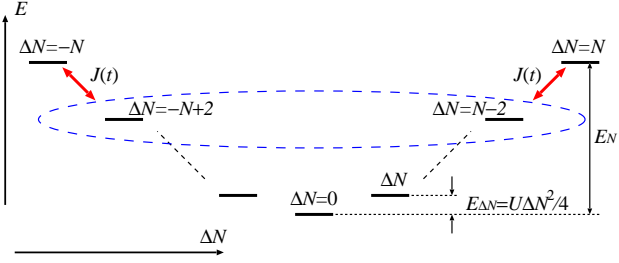}}}
\caption{\label{fig_leveldiagram}(Color online) Schematic diagram of
the present protocol for reducing the formation time of the NOON state.
}
%\end{center}
\end{figure}

Let us first evaluate the energy splitting $\Delta E_{\Delta N}$ of
the two degenerate states with the same value of $|\Delta N|\equiv
|N_R-N_L|$, where $N_R$ and $N_L$ are the numbers of particles in the
right and left wells, respectively.  The on-site interaction term,
which reads $\hat{H}_0\equiv U({\hat c}_R^\dagger {\hat c}_R^\dagger
{\hat c}_R {\hat c}_R +{\hat c}_L^\dagger {\hat c}_L^\dagger {\hat
c}_L {\hat c}_L)/2 = U\Delta \hat{N}^2/4 + U (\hat{N}^2 -2\hat{N})/4$,
gives the zeroth-order result for the energy as $E^{(0)}_{\Delta
N}\equiv U \Delta N^2/4$ (see Fig.\ \ref{fig_leveldiagram}).  Here
$\hat{N}\equiv {\hat c}_R^\dagger {\hat c}_R + {\hat c}_L^\dagger
{\hat c}_L$ and $\Delta\hat{N}\equiv {\hat c}_R^\dagger {\hat c}_R -
{\hat c}_L^\dagger {\hat c}_L$.  The splitting $\Delta E_{\Delta N}$
can be calculated by $|\Delta N|$th-order perturbation theory treating
the hopping term as a perturbation and we obtain
\begin{equation}
\Delta E_{\Delta N}=2JN \kappa^{-|\Delta N|+1}
\beta(N,\Delta N) ,
\label{deltae}
\end{equation}
with
\begin{equation}
\beta(N,\Delta N)
\equiv \frac{\left[\left(N+|\Delta N|\right)/2 \right]!\ N^{|\Delta N|-2} }
{\left[\left(N-|\Delta N|\right)/2 \right]!\ \left[(|\Delta N|-1)!\right]^2} .
\label{beta}
\end{equation}
The tunneling period $T_{\Delta N}$ between these states 
is given by $T_{\Delta N}=2\pi\hbar/\Delta E_{\Delta N}$.  Observe that, for
$\kappa \gg 1$, $T_{\Delta N}$ grows exponentially 
with $|\Delta N|$.

Our basic idea for reducing the formation time of the NOON state is to
modulate the tunneling matrix element we make a resonance between the
states of $|\Delta N|=N$ and $|\Delta N|=N-2$ (see Fig.\
\ref{fig_leveldiagram}), and take advantage of the tunneling between
the two states of $|\Delta N|=N-2$ whose period $T_{N-2}$ is much
shorter than $T_{N}$.  Using Eqs.\ (\ref{deltae}) and (\ref{beta}), we
estimate the reduction ratio as $T_N/T_{N-2} = \kappa^2
[(N-1)(N-2)]^2/N^3\sim \kappa^2 N$.  For, e.g., $N=5$ and $U/J=4$, the
reduction ratio is $T_N/T_{N-2}=460.8$.  From the general point of
view, this mechanism in which the higher-order co-tunneling is
assisted by the time-dependent tunneling matrix element would also be 
relevant to other contexts such as in Josephson qubits (e.g.,
\cite{review_josephson} for review).  Moreover, our theory directly
applies also to another situation: cold Bose gases in two hyperfine
states coupled by an external pumping field (see, e.g.,
\cite{micheli}).  In this case, $J$ corresponds to the Rabi frequency
of the external field and by controlling its intensity one can change
$J$.

We consider sinusoidal modulation of the tunneling matrix element and
employ the two-site Bose-Hubbard Hamiltonian (\ref{bh}) with $J$
replaced by $J(t)$ where
\begin{equation}
  J(t)\equiv J(1 + A \sin{\omega t}).
\label{jt}
\end{equation}
Here $A$ and $\omega$ are the amplitude and the frequency of the
modulation, respectively.  This modulation can be realized by
controlling the height of the barrier in the case of the double-well
potential (see, e.g., Ref.\ \cite{salmond} for reproducing $J(t)$ of
Eq.\ (\ref{jt}) in three-dimensional double-well potentials).  The
condition for a resonance between the states with $|\Delta N|=N$ and
$|\Delta N|=N-2$ is
\begin{equation}
  \hbar\omega \simeq E^{(0)}_{N} - E^{(0)}_{N-2} = U(N-1).
\label{res}
\end{equation}
As we shall see later, to obtain the periodic dynamics, the modulation
frequency must be much larger than the frequency of tunneling
oscillations, $\omega \gg 2\pi/T(\omega)$.

%\section{Results}

\begin{figure}[tbp]
%\begin{center}
\rotatebox{270}{
\resizebox{7.2cm}{!}
{\includegraphics{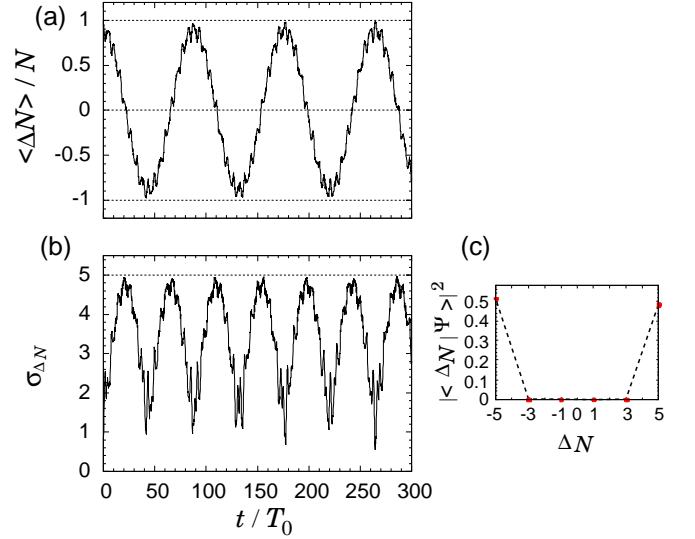}}}
\caption{\label{fig_dn}(Color online) Time evolution of 
$\langle\Delta N\rangle/N$ (a) and its fluctuation 
$\sigma_{\Delta N}$ (b)
for $N=5$, $U/J=4$, $A=0.1$, and $\hbar\omega/J=15$.  The tunneling
period is $T\simeq 88 T_0$.
Panel (c) shows the population of each component of $|\Delta N\rangle$ 
at $t=285.27 T_0$.
}
%\end{center}
\end{figure}

Now we calculate the time evolution using the two-site Bose-Hubbard
Hamiltonian with $J(t)$ of Eq.\ (\ref{jt}).  Here we take
$|N_L,N_R\rangle =|0,N\rangle$ as an initial condition.  In Fig.\
\ref{fig_dn}, we show the time evolution of $\langle\Delta N\rangle/N$
[panel (a)] and its fluctuation $\sigma_{\Delta N}\equiv\sqrt{\langle\Delta
N^2\rangle -\langle\Delta N\rangle^2}$ [panel (b)] for $N=5$, $U/J=4$,
$A=0.1$, and $\hbar\omega/J=15$ \cite{note_detuning} as an example.
Resulting dynamics has a periodic nature with a period $T\simeq 88
T_0$, which is much shorter than that without modulation evaluated by
Eqs.\ (\ref{deltae}) and (\ref{beta}): $T=1228.8 T_0$.  Here $T_0
\equiv 2\pi/\omega_0=\pi\hbar/J$, where $\omega_0\equiv 2J/\hbar$ and
$T_0$ are the oscillation frequency and the period of the tunneling
for $U=0$.  We also see that $\sigma_{\Delta N}$ almost reaches
$\sigma_{\Delta N}=N=5$ when $\langle\Delta N\rangle/N=0$
corresponding to a NOON state.  Figure \ref{fig_dn}(c) shows the
snapshot of the population of each component of $|\Delta N\rangle$ at
$t=285.27 T_0$; here, we have an almost perfect NOON state of
$\sigma_{\Delta N}=4.994$.

\begin{figure}[tbp]
%\begin{center}
\rotatebox{0}{
\resizebox{8.2cm}{!}
{\includegraphics{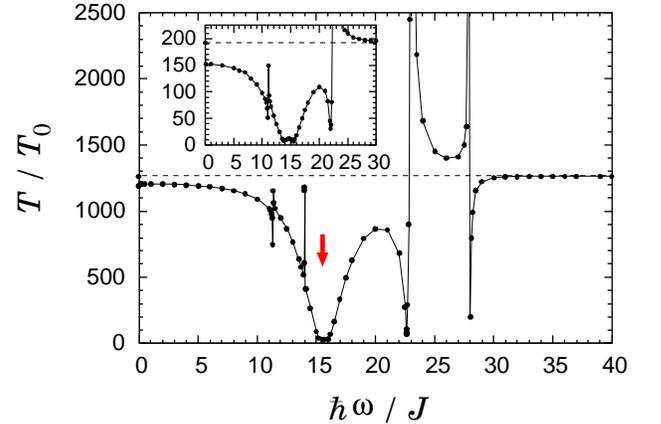}}}
\caption{\label{fig_tomega} (Color online)
The tunneling period $T$ as a function of $\omega$ for $N=5$, $U/J=4$,
and $A=0.1$.  There is a drastic reduction of $T$ in a wide range 
around $\hbar\omega/J\simeq 16$ (red arrow).  
The dashed lines show $T$ without modulation.
Very narrow resonances in the low-$\omega$ region of
$\hbar\omega/J\alt 10$ are not shown.
The inset shows for $N=4$, $U/J=5$, and $A=0.3$.
}
%\end{center}
\end{figure}

In Fig.\ \ref{fig_tomega}, we show the resulting tunneling period $T$
as a function of $\omega$ for the same parameters as in Fig.\
\ref{fig_dn}.  The resonance condition (\ref{res}) gives
$\hbar\omega/J=16$.  We note that there is a dramatic reduction of $T$
around $\hbar\omega/J=16$ with a wide width in $\omega$ (red
arrow) \cite{note_nowide}.  Let us first discuss the high- and low-$\omega$ 
regimes.  In the high-$\omega$ region of $\hbar\omega/J\agt
30$, tunneling period does not depend on $\omega$ any more and is
almost the same as in the case without modulation.  This is simply
because $\hbar\omega/J\agt 30$ is bigger than any other energy scales
in this problem and thus the system is insensible to the modulation of
$\hbar\omega/J\agt 30$.  In the low $\omega$ region of
$\hbar\omega/J\alt 10$ (but $\omega$ being much larger than the
frequency of the tunneling oscillation), $\omega$ dependence of $T$ is
weak (except for very narrow resonances, which are not shown in this
figure).  Since $2\pi/\omega\ll T$ even though $\omega$ is small, the
tunneling period can be evaluated by the time average of Eq.\
(\ref{deltae}) with replacing $J$ by $J(t)$ of Eq.\ (\ref{jt}):
$T_{\mbox{low $\omega$}}=2\pi/\langle\Delta E_{N, J(t)}\rangle_t$ with
$\langle\Delta E_{N,J(t)}\rangle_t \equiv \left\langle 2J(t)N
\left(J(t)/UN\right)^{N-1} \right\rangle_t \beta(N,N) =2JN
\kappa^{-N+1} \beta(N,N) \left\langle (1+A \sin{\omega t})^N
\right\rangle_t$.  This replacement of $J^N$ by $\langle
J^N(t)\rangle_t$ yields an effective increase of the energy splitting
and thus a reduction of $T$.

Next, we discuss the narrow resonances observed in Fig.\ \ref{fig_tomega}.
For this purpose, Floquet theory analysis 
is useful (e.g., Refs.\ \cite{salmond,floquet}).
The Floquet operator $\hat{F}$ is a mapping between the state 
at $t_0$ and the state after one modulation period at $t_0+2\pi/\omega$:
$|\Psi(t_0+2\pi/\omega)\rangle = \hat{F} |\Psi(t_0)\rangle$.
Here we get $\hat{F}$ as follows.
Starting from each state of the basis set $\{|\Delta N\rangle; 
\Delta N=-N, -N+2, ..., N\}$,
we follow the time evolution for one modulation period, $2\pi/\omega$.
Each resulting state forms a column of the matrix of $\hat{F}$.

\begin{figure}[tbp]
%\begin{center}
\rotatebox{0}{
\resizebox{7.2cm}{!}
{\includegraphics{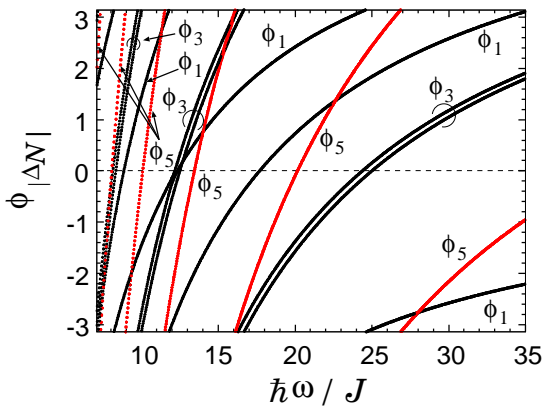}}}
\caption{\label{fig_floq}(Color online) 
Eigenvalues $\phi_{|\Delta N|}$ of the Floquet operator 
as functions of $\hbar\omega/J$ 
for the same parameters as in Fig.\ \ref{fig_tomega}: 
$N=5$, $U/J=4$, and $A=0.1$.
The red points (they look like lines except for the low-$\omega$ region) 
show $\phi_5$ and two sets of the red lines almost overlap.
Resonance occurs at $\omega$ where $\phi_5$ (red lines) crosses 
the other $\phi_{|\Delta N|}$; i.e., at
$\hbar\omega/J \simeq 11.3, 14.0, 15.5$-$16, 22.6$, and $28.0$.
}
%\end{center}
\end{figure}

In Fig.\ \ref{fig_floq}, we show the eigenvalues of $\hat{F}$ as
functions of $\hbar\omega/J$ for the same parameters as in Fig.\
\ref{fig_tomega}.  We denote the eigenvalue of $\hat{F}$ as
$\phi_{|\Delta N|}$ whose eigenstate has maximum amplitude at $\pm
\Delta N$ with even or odd parity in the Fock space labeled by $\Delta
N$ (However, there are significant populations in the other
components.  They become larger when two eigenvalues have a crossing
in Fig.\ \ref{fig_floq}.).  In this figure, the red points 
(they look like lines except for the low-$\omega$ region) show
$\phi_5$ and two sets of the red lines are so close that they appear
as a single line.  By comparing with Fig.\ \ref{fig_tomega}, we
observe that the resonances occur when $\phi_5$ (red lines) crosses
the other $\phi_{|\Delta N|}$ (black lines): the narrow resonances occur at
$\hbar\omega/J \simeq 11.3, 14.0, 22.6$, and $28.0$, where $\phi_5$
crosses $\phi_1$, and the wide one occurs at $\hbar\omega/J \simeq
15.5$-$16$, where $\phi_5$ almost simultaneously crosses two sets of
$\phi_3$.

\begin{figure}[tbp]
%\begin{center}
\rotatebox{0}{
\resizebox{8.2cm}{!}
{\includegraphics{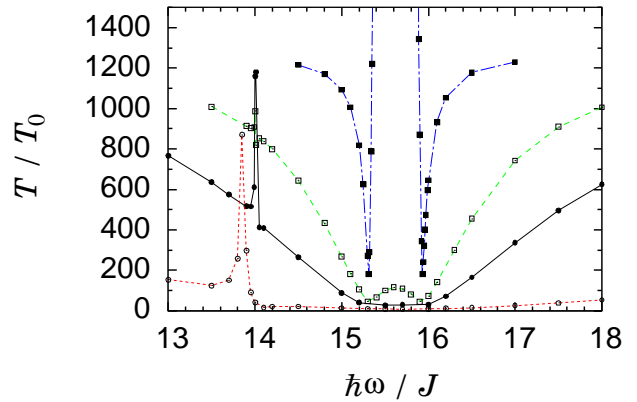}}}
\caption{\label{fig_width}(Color online) 
The tunneling period $T$ around $\hbar\omega/J=16$ as a function of
$\hbar\omega/J$ for various values of $A$.  The other parameters are
the same as in Figs.\ \ref{fig_dn} and \ref{fig_tomega}: $N=5$ and
$U/J=4$.  From the lower to the higher lines, $A=0.5$ 
(dotted line), $0.1$ (solid line), $0.05$ (dashed line), and $0.01$ 
(dashed-dotted line).
}
%\end{center}
\end{figure}

To understand the difference between the wide and narrow resonances,
we also show the tunneling period $T$ around $\hbar\omega/J=16$ for
various $A$ in Fig\ \ref{fig_width}.  With decreasing $A$, the
resonance becomes narrower and finally it separates into two narrow
resonances (see the case of $A=0.01$).  Note that the separation of
these two resonances is $\hbar\omega/J\simeq 0.75$, corresponding to
the energy splitting $\Delta E_{3}$ between the two states of $\Delta
N=\pm 3$.  This fact shows that, to obtain the wide resonance as in
the case of $A=0.5$ and 0.1 (and 0.05) in Fig.\ \ref{fig_width}, it is
necessary to couple the two states of $|\Delta N|=N-2$, whose energies
are different by $\Delta E_{N-2}$, to the states of $|\Delta N|=N$
(whose energy splitting is negligibly small).  Thus, the energy scale
of the amplitude of the modulation in the hopping term of Eq.\
(\ref{bh}) should be comparable to or larger than the energy splitting
of the states of $|\Delta N|=N-2$: $2JAN\agt \Delta E_{N-2}$.  From
Eqs.\ (\ref{deltae}) and (\ref{beta}), we then obtain $A \agt \Delta
E_{N-2}/2JN=N^{-1}(J/U)^{N-3} (N-1)(N-2)/(N-3)!$.  Concerning this
point, we have some remarks.  For $N=3$, this condition reads $A\agt
2/3$, which might be too large to be realized in a real double-well
potential.  For $N=2$, there is only one state of $\Delta N=N-2=0$,
and thus we never have a wide resonance.  On the other hand, for
larger $N$, further reduction of $T$ is possible by employing
additional modulations corresponding to wide resonances between the
states of $|\Delta N|=N-2$ and $N-4$, $N-4$ and $N-6$, etc., whose
energy splittings can be very small for larger $N$ [see Eq.\
(\ref{deltae})] \cite{note_twofreq}.  This ``scalability'' may be
useful for creating NOON states with larger $N$.

Note also that a tilt $\hat{V}_{\rm tilt}\equiv\Delta V \Delta\hat{N}/2$
of the double-well potential, which may be caused by the imperfection
of the double well or by the superimposed trapping, suppresses the
tunneling \cite{vavtunnel,tilt,note_tilt}.  To obtain the NOON state
using the present scheme, the tilt should be smaller than the energy
splitting for $|\Delta N|=N-2$: $\Delta V N \ll \Delta E_{N-2}$.

It is useful to know the width $\Gamma$ of the wide resonance, which
enables us to evaluate the formation time for a given $\omega$.  We
numerically obtain $\Gamma$ by the Lorentzian fit of the following
form: $T = (T_N-T_{N-2}) \left(1-\Gamma^2 [(\omega-\omega_0)^2 +
\Gamma^2]^{-1}\right) +T_{N-2}$.  This formula is constrained to
reproduce $T=T_{N-2}$ at $\omega=\omega_0$ \cite{note_omega0} and
$T=T_N$ in the limit of $\omega\rightarrow \infty$.  We find that
$\Gamma$ is almost proportional to $A$ and $U$, respectively.  From
the results of $N\alt 6$, we obtain the scaling relation as $\Gamma
\simeq 0.49 A\, U\, (N-1)(N-2)$.

%\section{Experimental feasibility}

Finally, we discuss the feasibility of creating larger NOON states in
current experiments \cite{mainzexp}.  In the present situation with
$N>2$, most serious decoherence process is due to three-body losses.
Let us estimate the decoherence time assuming the Gaussian wave
function in each well, $\psi({\bf r})=(N/\pi^{3/2}d_\perp^2 d_z)^{1/2}
\exp{\left[-(r_\perp^2/2d_\perp^2)-(z^2/2d_z^2)\right]}$, where we
take the direction of the double well in the $z$ direction and
$d_\perp$ and $d_z$ are the transverse and the longitudinal oscillator
lengths, respectively.  With the three-body rate constant $K_3$, the decoherence
time $\tau_3$ is given by $\tau_3^{-1} = K_3 \int d^3r\ |\psi|^6 =
(\sqrt{3}\pi)^{-3}K_3 N^3 (d_\perp^4 d_z^2)^{-1}\simeq (4\pi/3)^3 K_3
N^3 s_\perp s_z^{1/2} (\lambda_\perp^2 \lambda_z^4)^{-1}$.  Here
$\lambda_z$ and $\lambda_\perp$ are the wavelengths of the lasers in
the longitudinal and transverse directions, respectively, 
and $s_\perp$ and $s_z$ are
the longitudinal and transverse lattice heights, respectively, in units of
the recoil energy in the longitudinal direction \cite{note_ex}.

In the experiment of Ref.\ \cite{mainzexp}, $\lambda_z=765$ nm,
$\lambda_\perp=843$ nm, $s_z\simeq 10$ (for $U/J=5$), $s_\perp=33$,
and $K_3=5.8\times 10^{-42}$ Hz m$^6$ for $^{87}$Rb \cite{k3rb}, and
we obtain $\tau_3\simeq 16$ ms for $N=4$.  Using the present scheme, the
tunneling period for $N=4$ and $U/J=5$ \cite{note_condtilt} can be
reduced to $\alt 20 T_0$ (see inset of Fig.\ \ref{fig_tomega}),
corresponding to $\alt 14.4$ ms in this experiment.  Thus, the system
safely undergoes one period of the oscillation in the decoherence
time.  By the same technique used in Ref.\ \cite{mainzexp}, one can
measure $\langle \Delta N\rangle$.  Its oscillatory behavior is a firm
evidence for the coherency of the system and rules out the possibility
of the 50:50 mixture of $|N, 0\rangle$ and $|0, N\rangle$.  From the
oscillation period of $\langle \Delta N\rangle$, one can also reject
the possibility of $|N/2,N/2\rangle$ at $\langle \Delta N\rangle=0$.
The combination of the above two evidences establishes the realization of
a NOON state.

%\section{Conclusion}

In conclusion, we have found the higher-order co-tunneling can be
assisted by sinusoidally modulating the tunneling matrix element.
Using this mechanism, we have proposed an efficient scheme for
periodically creating almost perfect NOON states in two-state Bose
systems.  This scheme is scalable in a sense that, for larger number
of particles, further reduction is possible using more than one
frequency.  With this scheme, a NOON state of four particles can be
realized in current experiments of cold Bose gases in a double-well
potential.

\medskip
%\begin{acknowledgements}
The author is grateful to Augusto Smerzi and Chris Pethick. 
He also thanks L. D. Carr, P. Hyllus, F. Piazza, 
W. P. Reinhardt, and M. Ueda.
%\end{acknowledgements}


\begin{thebibliography}{99}
\bibitem{note_map} In Ref.\ \cite{vavtunnel}, we pointed out
that there is a one-to-one correspondence between the
vortex-antivortex states discussed there and the Fock 
states with all particles in the right or left wells in the
double-well potential.
\bibitem{note_anharm} While this description is within the two-mode
approximation, such a situation could be realized by setting the level spacing
in each well much larger than the interaction energy per particle.
\bibitem{note_detuning}
To obtain an almost perfect NOON state, one should slightly detune
$\omega$ from Eq.\ (\ref{res}).
\bibitem{note_twofreq} 
We set $J(t)$ as, e.g.,
$J(t)=J[1+A(\sin{\omega_1 t}+\sin{\omega_2 t})]$,
with $\omega_1$ and $\omega_2$ corresponding to transitions between
$|\Delta N|=N$ and $N-2$, and between $N-2$ and $N-4$, respectively.
\bibitem{note_omega0}
Since the central value of $\omega$ of the wide resonance is slightly 
lower than that given by Eq.\ (\ref{res}),
we adjust $\omega_0$ in the fitting.
\bibitem{note_ex}
To avoid excitations to higher energy levels 
by the modulation, $\hbar\omega$ should be much smaller than 
the level spacing in the well.
For simplicity we consider $d_\perp=d_z ( \equiv d_w=\sqrt{\hbar/m\omega_w})$,
where $\omega_w$ is the frequency of each well.
Since $\hbar\omega\sim UN$ from Eq.\ (\ref{res}), 
and $UN/\hbar\omega_w\simeq \sqrt{2/\pi}(a_{\rm s}/d_w) N$ \cite{smerzi}, 
where $a_{\rm s}$ is the $s$-wave scattering length,
that condition reads
$Na_{\rm s}/d_w \ll 1$, which is not difficult to satisfy for $N=O(1)$.
\bibitem{note_condtilt} For these parameters, the condition for the tilt, 
$\Delta V N \ll \Delta E_{N-2}$, reads $\Delta V/J \ll 0.6$, which can be
easily satisfied.
\end{thebibliography}
\end{document}